\documentclass[9pt,twocolumn,twoside]{pnas-new}
\usepackage{amssymb}
\usepackage{amsmath,bm}
\usepackage{graphicx,color}
\usepackage{bbm}
\usepackage{multirow}
\templatetype{pnasresearcharticle} % Choose template
\title{Shear bands as manifestation of a criticality in yielding amorphous solids}

\author[a,]{Giorgio Parisi}
\author[b,1]{Itamar Procaccia}
\author[b]{Corrado Rainone}
\author[b]{Murari Singh}

\affil[a]{Dipartimento di Fisica, Sapienza Universit\'a di Roma, INFN, Sezione di Roma I, IPFC -- CNR, Piazzale Aldo Moro 2, I-00185 Roma, Italy}
\affil[b]{Department of Chemical Physics, the Weizmann Institute of Science, Rehovot 76100, Israel}

\leadauthor{Parisi}

\significancestatement{The art of making structural, polymeric, and metallic glasses is rapidly developing with many applications. A limitation is that under increasing external strain all amorphous solids have a yield stress which, when exceeded, results in a plastic response leading to mechanical failure. Understanding this is crucial for assessing the risk of failure of glassy materials under loads. The universality of the mechanical yield requires a theory that is general enough to transcend the microscopic details of different glasses, which all show similar stress-strain curves with a yield point. We provide what appears to be the first general theory which is thermodynamic in nature, showing that the mechanical yield is a spinodal criticality in an appropriately constructed free energy landscape.}

\authorcontributions{GP, IP and CR designed research. IP, CR and MS performed research and analyzed data, and IP and CR wrote the paper.}
\authordeclaration{The authors declare no conflict of interest.}
\correspondingauthor{\textsuperscript{1}To whom correspondence should be addressed. E-mail: Itamar.Procaccia@gmail.com}


\begin{abstract}
Amorphous solids increase their stress as a function of an applied strain until a mechanical yield point
whereupon the stress cannot increase anymore, afterwards exhibiting a steady state with a constant mean stress. In stress
controlled experiments the system simply breaks when pushed beyond this mean stress. The ubiquity of this
phenomenon over a huge variety of amorphous solids calls for a generic theory that is free of microscopic
details. Here we offer such a theory: the mechanical yield is a thermodynamic phase
transition, where yield occurs as a spinodal phenomenon. At the spinodal point there exists a divergent
correlation length which is associated with the system-spanning instabilities (known also as shear
bands) which are typical to the mechanical yield. The theory, the order parameter used and the correlation functions which exhibit the divergent correlation length are universal in nature and can be applied
to any amorphous solids that undergo mechanical yield.
\end{abstract}

\dates{This manuscript was compiled on \today}
\doi{\url{www.pnas.org/cgi/doi/10.1073/pnas.XXXXXXXXXX}}

\begin{document}

% Optional adjustment to line up main text (after abstract) of first page with line numbers, when using both lineno and twocolumn options.
% You should only change this length when you've finalised the article contents.
\verticaladjustment{-2pt}

\maketitle
\thispagestyle{firststyle}
\ifthenelse{\boolean{shortarticle}}{\ifthenelse{\boolean{singlecolumn}}{\abscontentformatted}{\abscontent}}{}

% If your first paragraph (i.e. with the \dropcap) contains a list environment (quote, quotation, theorem, definition, enumerate, itemize...), the line after the list may have some extra indentation. If this is the case, add \parshape=0 to the end of the list environment.
\dropcap{A} solid, be it crystalline or amorphous, is operatively defined as any material capable to respond elastically to an externally applied shear deformation~\cite{59LL}. However, any solid material, when subject to a large enough shear-strain, finally undergoes a mechanical yield. Here we focus on the mechanical yield of amorphous materials such as molecular and colloidal glasses, foams, and granular matter. The phenomenology exhibited by the yielding point within this vast class of materials, as reported in countless strain-controlled simulations~\cite{04VBB,04ML,05DA,06TLB,06LM,09LP,11RTV} and experiments~\cite{06SLG,13KTG,13NSSMM} shows a remarkable degree of universality despite the highly varied nature of the model systems involved. Among these universal features is the presence, at the onset of flow at yielding, of system spanning excitations referred to as shear-bands~\cite{11BB,12DHP}, wherein the shear strongly localizes, leaving the rest of the material unperturbed. This phenomenon is of capital importance for engineering applications as it is responsible for the brittleness typical of glassy materials, in particular metallic glasses~\cite{06AG}, whose potential for practical use is stymied by their tendency to shear-band and fracture~\cite{12DHP,13DHP,13DGMPS}.

In athermal amorphous solids the phenomenon has universal features. For strains $\gamma$ smaller than some critical value denoted as $\gamma_{\rm Y}$ the stress in the material grows on the average when the strain is increased. After yield the stress cannot grow on the average, no matter how much the strain is increased. The universality of the basic phenomenology of yielding begs a picture of its characteristics in terms of a universal theory, in the sense that such a theory should rely on a statistical-mechanical framework and be independent of details such as chemical composition and production process of the material. This need was addressed in a recent work~\cite{16JPRS}, wherein building up from ideas first advanced in \cite{16RU} there emerged a picture of mechanical yielding as a first-order phenomenon, i.e. as a discontinuous phase transition in a suitable overlap order parameter $Q_{ab}$ (defined in Eq.~(\ref{defQ}) below) which jumps from a value of order 1 to a value of order zero as strain is increased above the yielding threshold $\gamma_Y$. The physical meaning of this observation is that before yielding the amorphous system was limited to a small patch in the configuration space, very far from any
kind of ergodicity. The yielding transition is an opening of a much larger available configuration space, whereupon the system is ergodized subject to the constraint of constant mean stress. Within this framework, the yielding transition is essentially envisioned as a spinodal point~\cite{16UZ} i.e. the point where the metastable, high $Q_{ab}$ glassy patch of available configurations becomes unstable with respect to a new phase with low $Q_{ab}$, associated with an ergodized system in the presence of disorder~\cite{16NBT}. A paradigmatic example of such a spinodal is the Mode Coupling crossover~\cite{11BB}, characterized by dynamical slowing down and heterogeneities, whose behavior is characterized by a \emph{dynamical lengthscale} which can be extracted from suitable multi-point correlators~\cite{11BB}. According to our picture, this kind of critical behavior should also be found at the yielding transition, conditional that one is able to derive the expression of the right correlator to measure. This suggestion seems even more reasonable in light of a recent study~\cite{16SCH} wherein the similarity of shear bands with dynamical heterogeneities has been pointed out; also, some oscillatory shear simulations seem to indicate that a slow-down of the dynamics on approaching yielding may indeed be present~\cite{17RL,LAS16}. It is important to stress here that the reason that a spinodal point can be exposed and measured is that the
glassy time scales and the athermal conditions stabilize the metastable system until the spinodal point is crossed and the system becomes unstable against constrained ergodization.

Within a generic statistical-mechanical theory, formulated in terms of a suitable Gibbs free energy $G[\phi]$ (i.e. the free-energy for fixed order parameter $\phi$), stable phases are identified with its points of minimum in $\phi$, and phase transitions happen when the curvature of these minima goes to zero, inducing a critical behavior which manifests diverging susceptibilities-fluctuations, critical slowing down of the dynamics, and growing correlation lengths~\cite{02Z-J}. At a spinodal point, for example, one such minimum becomes unstable and transforms into a saddle. In the case of the order parameter $Q_{ab}$ the general form of the free energy $s[Q_{ab}]$ had been already derived and studied (see~\cite{98DKT} for a review) in the context of the theory of replicas originally developed for the study of spin-glasses, and its properties, at least at mean-field level, are well known (we refer to~\cite{15RUYZ,16RU} for the derivation of $s[Q_{ab}]$ in the specific case of mean-field hard spheres); the matrix of second derivatives (or, using a more field-theoretic terminology, the mass matrix) is not diagonal in the base of $Q_{ab}$, and after diagonalization is found to have only three distinct modes, or masses~\cite{98DKT}. Of these, the most relevant ones are the so called \emph{replicon mode} $\lambda_R$, which for example goes to zero at the newly proposed Gardner transition~\cite{14CKPUZ}, and the \emph{longitudinal mode} $\lambda_L$ which is instead related to spinodal points \cite{16RU,16UZ} such as our yielding transition.  In the Supplementary Information to this paper we review briefly the background theory
that is at the basis of the present approach.

In this paper, we build up from the results of~\cite{16JPRS} and, following the line of reasoning formulated above, we employ the expression of the correlation function relative to the longitudinal mode $\lambda_L$ as it can be derived from the replicated field theory~\cite{98DKT} to reveal the critical features of the yielding transition. We measure this correlator in numerical simulation, and use it to expose the critical properties of the yielding transition, showing how shear bands manifest the diverging correlation length encoded in this correlator. We show how the order parameter $Q_{ab}$ and its associated replicated field theory are thereby able to provide a unified and universal picture of the yielding transition in terms of a spinodal point in presence of disorder, with an associated criticality.

\section*{The correlation functions\label{sec:def}}

The relevant order parameter for the problem at hand is the overlap function $Q_{ab}$ which
measures the distance between two configurations ``$a$" and ``$b$" of the same system. Denoting the
position of the $i$th particle as $r_i^a$ in configuration ``$a$" and $r_i^b$ in configuration ``$b$"
we define
\begin{equation}
  Q_{ab} \equiv \frac{1}{N}\sum_{i=1}^N\theta(\ell-|{\bf r}_i^a-{\bf r}_i^b|)\ ,
  \label{defQ}
 \end{equation}
 where $\theta(x)$ is the Heaviside step function and $\ell$ is a constant length which is taken
 below to be 1/3 in Lennard-Jones units (see below for numerical details). Thus $Q_{ab}=1$ for two
 identical configurations and $Q_{ab}=0$ when the distance between the positions of all the particles
 $i$ in the two configurations exceed $\ell$.
\
 Based on the introductory discussion, we now derive an expression for the correlator associated with the longitudinal mode, from whence one can extract the correlation length associated with the onset of criticality at the yielding point, and define an associated susceptibility which will shoot up as the yielding point is approached. The first step is to ``localize" the overlap function and define the ${\bf r}$-dependent quantity
 \begin{equation}
  Q_{ab}({\bf r}) \equiv \sum_{i=1}^N\theta(\ell-|{\bf r}_i^a-{\bf r}_i^b|)\delta ({\bf r} -{\bf r}_i^a)\ ,
  \label{defQr}
 \end{equation}
Next, as mentioned above, the expression for the longitudinal correlator in terms of four-replica correlation functions can be found by diagonalization of the correlation matrix $G_{ab;cd}$, which is defined as the inverse of the mass matrix $M_{ab;cd}$ of the replicated field theory of the overlap order parameter $Q_{ab}$~\cite{98DKT}. The derivation is a matter of standard diagonalization algebra, so we shall not report it here and refer to the SI for the details. The expression, employed for example in~\cite{14BB1,14BB2} in the case of a model with spins on a lattice, reads for athermal systems
 \begin{equation}
  G_L({\bf r}) = 2G_R({\bf r}) - \Gamma_2({\bf r}),
 \end{equation}
 with the definitions
 \begin{eqnarray}
 G_R({\bf r}) &\equiv& \overline{\left<Q_{ab}(r)Q_{ab}(0)\right>} - 2\overline{\left<Q_{ab}(r)Q_{ac}(0)\right>}\\
 &&+ \overline{ \left<Q_{ab}(r)\right>\left<Q_{cd}(0)\right>},\nonumber\\
 \Gamma_2({\bf r}) &\equiv& \overline{\left<Q_{ab}({\bf r})Q_{ab}(0)\right> - \left<Q_{ab}({\bf r})\right>\left<Q_{ab}(0)\right>}\ .
 \end{eqnarray}
Here angular brackets denote a thermal average in the thermal case and an evaluation in an inherent state
in the athermal case; an $\overline{(\bullet)}$ indicates an average over different samples of the glass. The quantity $G_R({\bf r})$ is the correlation function of the replicon mode~\cite{98DKT} and $\Gamma_2({\bf r})$ is just the garden-variety four-point correlator.

 Using these definitions and taking Eq. \eqref{defQr} into account, the quantities we compute in numerical simulation, before taking the ensemble average, are (see the SI and Ref.~\cite{16BCJPSZ}):
 \begin{equation}
 \tilde \Gamma_2({\bf r}) = \frac{ \sum_{i\neq j}(u^{ab}_i-Q_{ab}) (u^{ab}_j-Q_{ab})\delta({\bf r}-({\bf r}_{i}^a-{\bf r}_{j}^a)) }{ \sum_{i\neq j}\delta({\bf r}-({\bf r}_{i}^a-{\bf r}_{j}^a)) }
 \end{equation}
 and
 \begin{equation}
  \tilde G_R({\bf r})  = \frac{ \sum_{i\neq j}[u_i^{ab}u_j^{ab} - 2u_i^{ab}u_j^{ac} + Q_{ab} ~Q_{cd}]\delta({\bf r}-({\bf r}_{i}^a-{\bf r}_{j}^a)) }{ \sum_{i\neq j}\delta({\bf r}-({\bf r}_{i}^a-{\bf r}_{j}^a)) }.
 \end{equation}
 with
 \begin{equation}
  u^{ab}_i \equiv \theta(\ell-|{\bf r}_i^a-{\bf r}_i^b|) \ .
 \end{equation}
These four-replica objects can be computed for any quadruplet of distinct replicas. The ensemble averaged correlation functions are simply obtained as $\Gamma_2\equiv \overline{\tilde \Gamma^{ab}_2}$ and $G_R\equiv \overline{\tilde G^{ab}_R}$, and cf. the SI for a proof. We stress that one must keep the full space dependence of the correlators in the definitions above, as the introduction of shear breaks the rotational symmetry of the glass samples and so the correlators are not just functions of a distance $r$.

\section*{Numerics}

To measure the quantities defined above, we performed molecular dynamics simulations of a Kob-Andersen 65-35\% Lennard Jones Binary Mixture in $2d$. We have three system sizes, $N=1000$, $N=4000$ and $N=10000$. We chose $Q_{12}$ with $\ell = 0.3$ in LJ units, but verified that changes in $\ell$ leave the emerging picture invariant.

Following the procedure reported in Ref.~\cite{16JPRS}, as a first step we prepared a glass by equilibrating the system at $T=0.4$, and then quenching it (the rate is $10^{-6}$) down to $T=1\cdot10^{-6}$ into a glassy configuration. The sample is then heated up again to $T=0.2$, and a starting configuration of particle positions is chosen at this temperature. Note that while at $T=0.4$ equilibration is sufficiently fast, at $T=0.2$ the computation time is much shorter than the relaxation time. The configuration is then assigned a set of velocities randomly drawn from the Maxwell-Boltzmann distribution at $T=0.2$, and these different samples are then quenched down to $T=0$ at a rate of $0.1$. This procedure can be repeated any number of times (say 100 times), and it allows us to get a sampling of the configurations, or replicas, inside one single ``glassy patch". We then perform this procedure again, using each time a different configuration from the parent melt at $T=0.4$, and in doing so we get an ensemble of these glassy patches, each of them representing a distinct glass sample. For each of these patches, we measure the four-replica correlators defined above for any distinct quadruplet of replicas, averaging the result over any possible permutations of the quadruplet to gain statistics~\cite{14BB2}. The ensemble average is then performed by averaging the result over all the glass samples. To perform these measurement, below we use 100 patches for $N=1000$, each with 100 configurations, 100 patches for $N=4000$  each with 50 configurations and 50 patches for
$N=10000$ each with 50 configurations.
A strain $\gamma_{xy}$ (denoted below as $\gamma$) is then applied quasi-statically to all configurations in all patches. In this protocol after every
step of increased strain the system undergoes energy gradient minimization to return to mechanical equilibrium. This creates an ensemble of \emph{strained patches} for every value of the strain parameter $\gamma$, from whence we measure again the above defined correlators, which then become functions of the strain $\gamma$. This is simply a consequence of the response of the configurations, i.e. each position ${\bf r}_i$ in the definitions above becomes ${\bf r}_i(\gamma)$. Thus for example $G_R({\bf r})$ becomes $G_R({\bf r}; \gamma)$ etc. We are interested in the behavior of the correlators as the yielding point $\gamma_{\rm Y}$ is approached.
%Having 100 replicas allows us to compute $Q_{12}$ $100\times 99/2$ times,
%and averaging over these many pairs we obtain $\overline{Q_{12}}$. Finally we repeat the whole
%procedure many times (say 100 times) to obtain a probability distribution of $\overline{Q_{12}}$ that is
%denoted below as $P(\overline{Q_{12}})$. When we strain the system we compute $\overline{Q_{12}(\gamma)}$
%in which we compare all the pairs of replicas at the same value of $\gamma$.
%%%%%%%%%%%%%%%%%%%%%%%%%%%%%%%%%%%%%
\begin{figure}[h!]
 \includegraphics[width=0.5\textwidth]{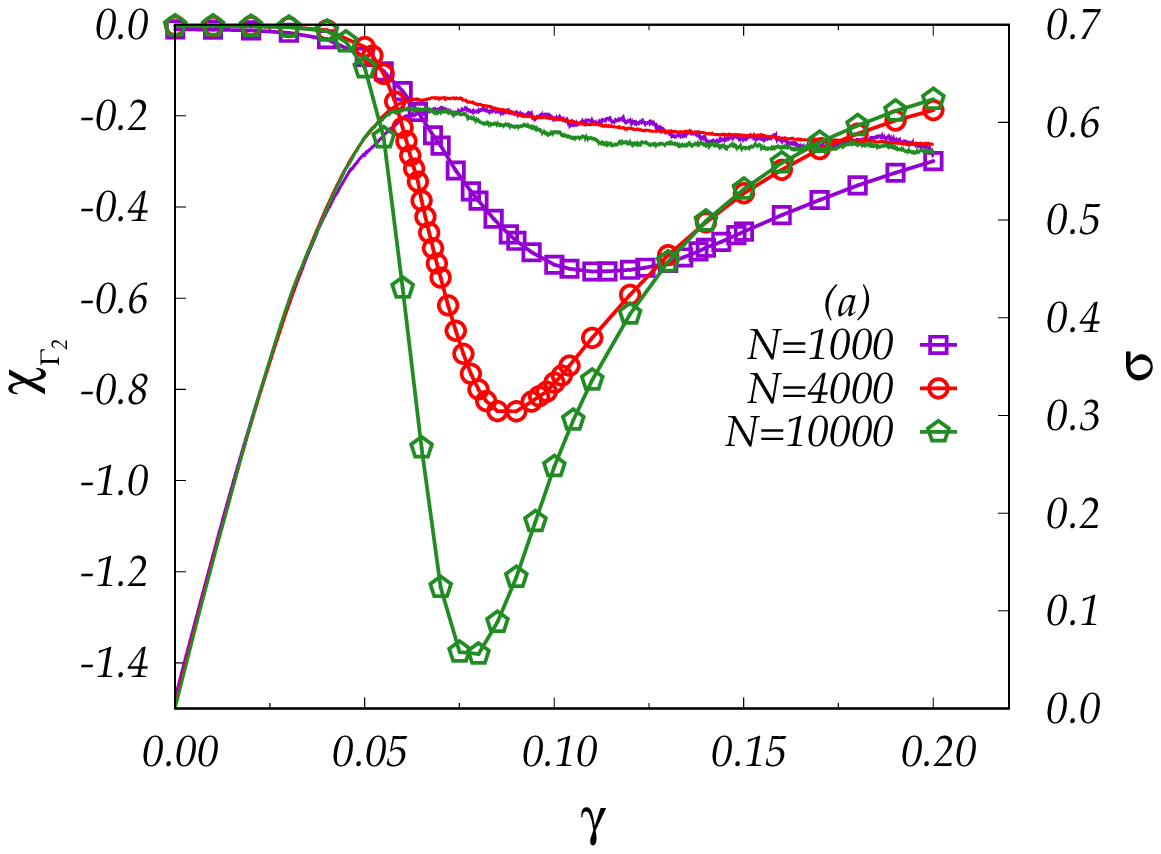}
 \includegraphics[width=0.5\textwidth]{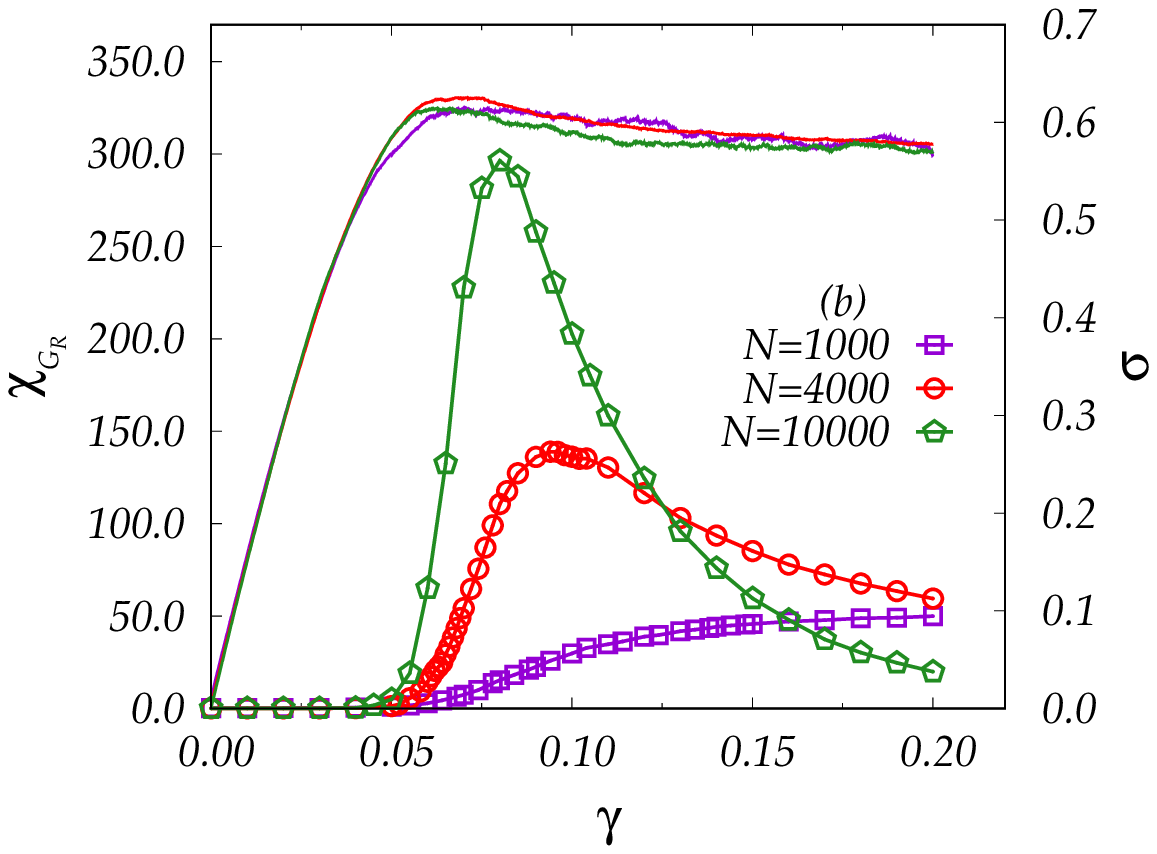}
 \caption{The susceptibilities $\chi_{_{\Gamma_2}}$ (upper panel) and $\chi_{_{G_R}}$ (lower panel) as a function of $\gamma$ for the three systems sizes available. Superimposed are the stress vs. strain curves for comparison.
 The color code is violet for $N=1000$, red for 4000 and green for 10000.}  \label{sus}
\end{figure}
%%%%%%%%%%%%%%%%%%%%%%%%%%%%%%%%%%%%%%%%%%%%%%%%%%%%%%%%%%%%

\section*{Results}

We consider first the susceptibilities $\chi_{_{G_L}}, \chi_{_{G_R}}$ and $\chi_{_{\Gamma_2}}$ that can be obtained from the correlators, for example
\begin{equation}
 \chi_{_{G_L}}(\gamma) \equiv \int d^2 x\ G_L(x,y;\gamma) \ .
\end{equation}
 In figure \ref{sus} upper panel we show the susceptibility $\chi_{_{\Gamma_2}}$ as a function of $\gamma$ for the three system sizes at our disposal. Superimposed are the stress vs. strain curves obtained by averaging
 the individual curves over all the available configurations and glass samples. One sees very clearly the singularity that
 develops near the yield point as a function of the system size. In the lower panel of the same figure we show the susceptibility $\chi_{_{G_R}}$ as a function of the strain $\gamma$, again with the stress-strain curve superimposed for comparison. As we expected, the susceptibilities show a distinct peak at the spinodal point $\gamma_Y$ wherein yielding occurs. Since $\chi_{_{\Gamma_2}}$ is much smaller in amplitude than $\chi_{_{G_R}}$ there is no much new information in $\chi_{_{G_L}}$ which is approximately 2$\chi_{_{G_R}}$.

\begin{figure}[htb!]
 \includegraphics[width=0.5\textwidth]{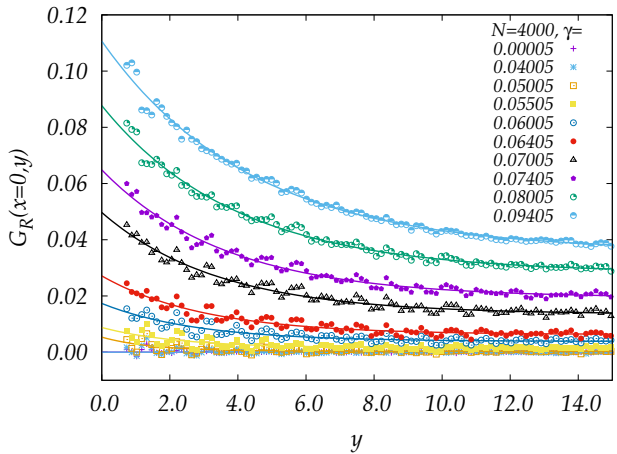}
 \caption{The function $G_R(x=0,y;\gamma)$ for various values of $\gamma$ from $5\times 10^{-5}$ to 0.09405.  Note the increase in the
 overall amplitude of the correlator as well as the increase in the correlation length. The lines through
 the data are the fit function Eq.~(\ref{fit}).}
 \label{cut}
\end{figure}

\begin{figure}[h!]
\includegraphics[width=0.5\textwidth]{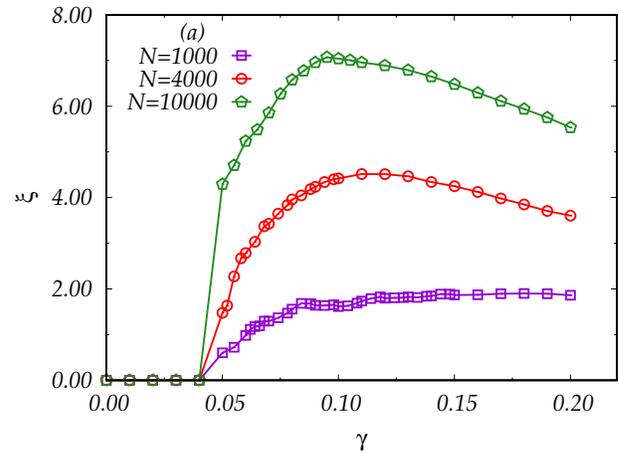}
\includegraphics[width=0.5\textwidth]{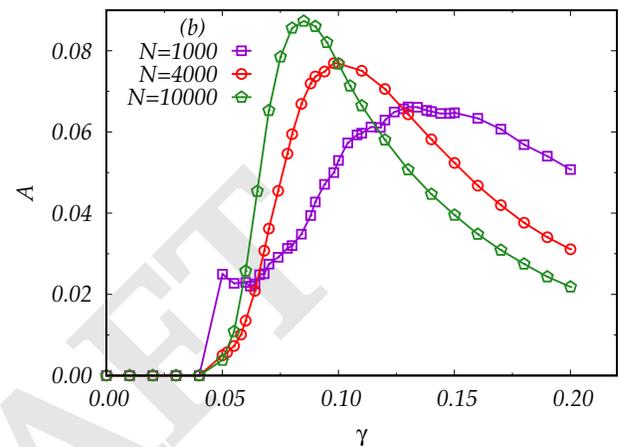}
\includegraphics[width=0.5\textwidth]{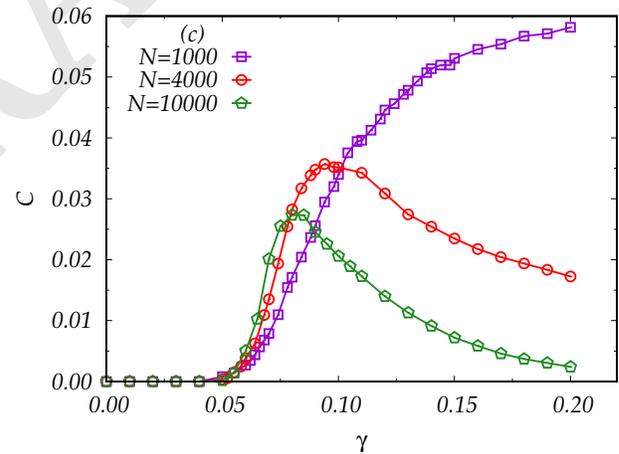}
 \caption{The $\gamma$ dependence of the correlation length $\xi(\gamma)$, the amplitude $A(\gamma)$ and the constant $C(\gamma)$ in the
 best fit to the function $G_R(x=0,y;\gamma)$, cf. Eq.~\ref{fit}. }
 \label{results}
\end{figure}

 More detailed information is provided by the full dependence of the correlators on their arguments.
 To see most clearly the change in the correlators as the spinodal point is approached, it is best
 to consider for example the one-dimensional function $G_R(x=0,y;\gamma)$, shown for $N=4000$ in Fig.~\ref{cut}.
%%%%%%%%%%%%%%%%%%%%%%%%%%%%%%%%%%%%%%%%%%%%%%%%%
%%%%%%%%%%%%%%%%%%%%%%%%%%%%%%%%%%%%%%%%%%%%%%%%%%%%%%%%%%%%%%%%%%%%%%%%%%%
Similar results for the other systems sizes are available in the SI. We note that the correlator changes both in amplitude and in extent when we approach the critical point.
To quantify these changes we fit a 3 parameter function to $G_R(x=0,y)$ in the form
\begin{equation}
G_R(x=0,y;\gamma)\approx C+ A \exp(-y/\xi) \ ,
\label{fit}
\end{equation}
where all the fitting coefficients are functions of $\gamma$.
In Fig.~\ref{results} we present the $\gamma$ dependence of the amplitude $A(\gamma)$, the constant $C(\gamma)$ and the correlation length $\xi(\gamma)$.

It is interesting to notice that the constant $C$ decreases with the system size, presumably becoming irrelevant
in the thermodynamic limit. The amplitude $A$ is still increasing with the system size, and it is difficult
to assert whether it converges or not. On the other hand we can safely conclude that the data present a strong evidence for the increase in the correlation length; it is very likely that it should diverge in the
thermodynamic limit.

A relevant question is whether one can define critical exponents that can be measured also in
experimental situations, and whether such exponents can be computed from theory, even on the
mean-field level. Clearly, the standard thermal mean-field approach cannot be employed, since
averages here are computed over replicas at $T=0$, and fluctuations due to quenched disorder are expected
to dominate the thermal fluctuations that stem only from the mother super-cooled liquid from which
the replicas at $T=0$ are created. Considerations of the effect of such fluctuations are beyond
the scope of this paper and will be discussed elsewhere.

\section*{Physical interpretation}

\begin{figure}[h!]
\includegraphics[width=0.5\textwidth]{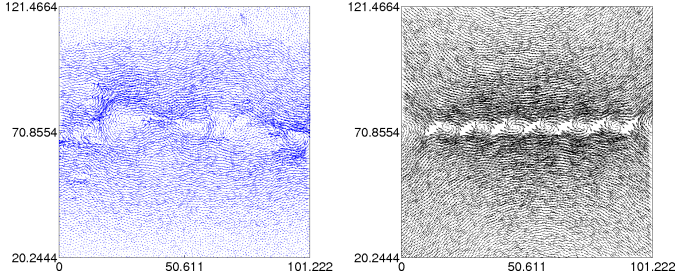}
\caption{Example of the spontaneous plastic event exhibiting a concatenation of a series of Eshelby quadrupoles resulting in a correlated displacement field with shear localization over a thin region which is
system spanning. Left panel: direct numerical simulations. Right panel: inserted line of Eshelby quadrupoles.}
\label{shearband}
\end{figure}

To conclude this paper we present a physical interpretation to these new insights, connecting
them to what is known about the mechanical yield in athermal amorphous solids. The most important
characteristic of the mechanical yield in athermal amorphous solids is the change from plastic
responses that are localized, typically in the form of Eshelby quadrupoles, to subextensive plastic
events that are system spanning \cite{10KLP,15HJPS}. The energy drops associated with the localized Eshelby quadrupoles are system
size independent, scaling like $N^0$ where $N$ is the total number of particles in the system.
Mechanical yield is associated with the spontaneous appearance of concatenated lines of quadrupoles
(in 2 dimensions, or planes in 3 dimensions, \cite{12DHP,13DHP,13DGMPS}). The latter are associated with energy drops that are subextensive, scaling like $N^{1/3}$ in 2 dimensions. Importantly, the concatenated lines of quadrupoles change drastically the displacement field associated with the plastic events. Each quadrupole has an arm with a displacement field pointing outward and an arm with the displacement field pointing inward.
When the quadrupole is isolated the displacement field decays algebraically to infinity. In contrast,
when the quadrupoles are organized in the line there is a global connection between the outgoing direction
of one quadrupole and the incoming direction of the next, making the displacement field strongly localized
around the line of quadrupoles (or around a plane in 3 dimensions), and all the shear is there. This is a microscopic shear band. An example of the displacement field associated with such as system spanning
event is shown in Fig.~\ref{shearband}, and see Ref.~\cite{12DHP} for details. The
main point of this paper is that the highly correlated phenomenon of such a shear band can only occur
when there exists a correlation length that approaches the system size in magnitude. This is the correlation
length $\xi$ that is identified in this paper, and cf. the upper panel of Fig.~\ref{results}.
%%%%%%%%%%%%%%%%%%%%%%%%%%%%%%%%%%%%%%%%%%%%
%%%%%%%%%%%%%%%%%%%%%%%%%%%%%%%%%%%%%%%%%

To understand the relevance of the spinodal point for this scenario, we provide two figures that
were obtained in Ref.~\cite{16JPRS}.
\begin{figure}[h!]
\begin{center}
\includegraphics[width=0.45\textwidth]{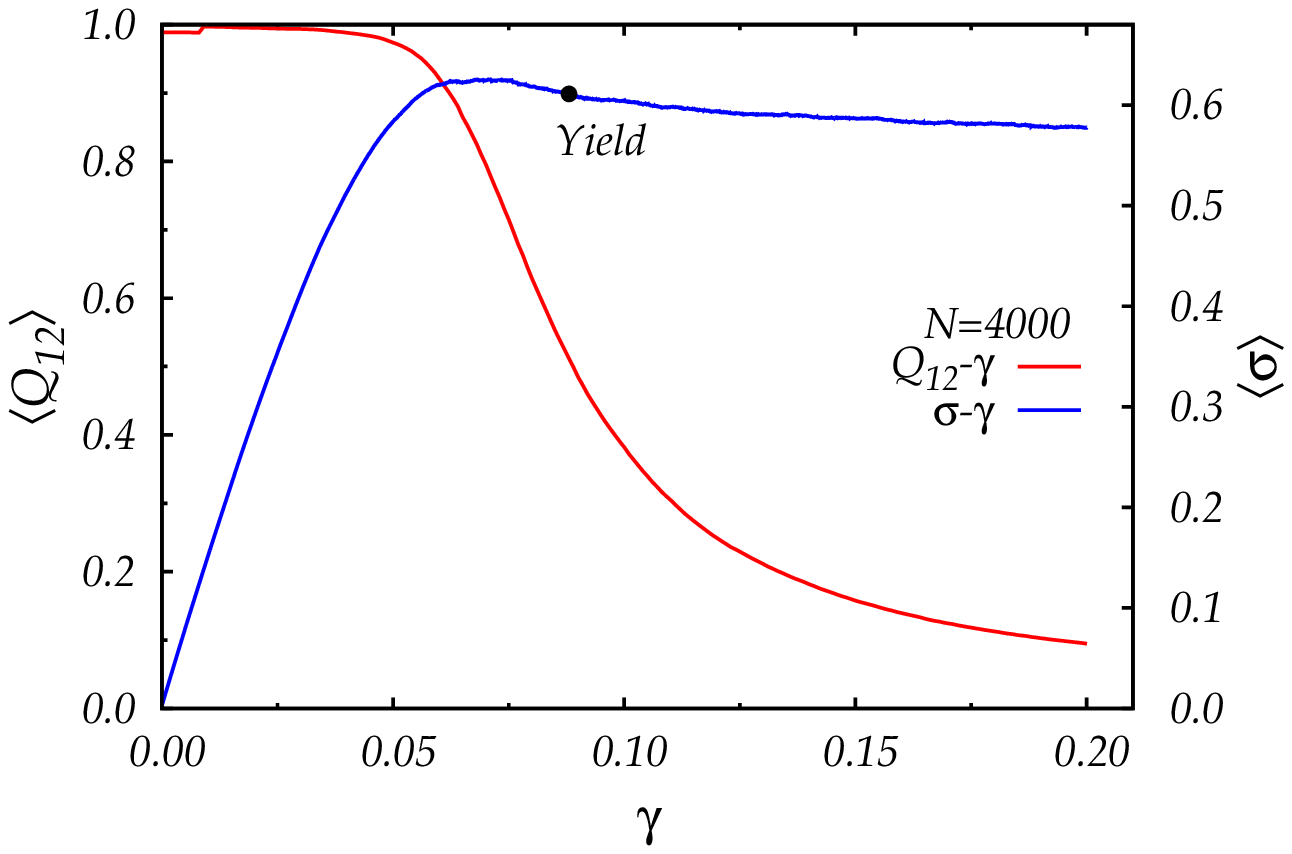}
  \includegraphics[width=0.45\textwidth]{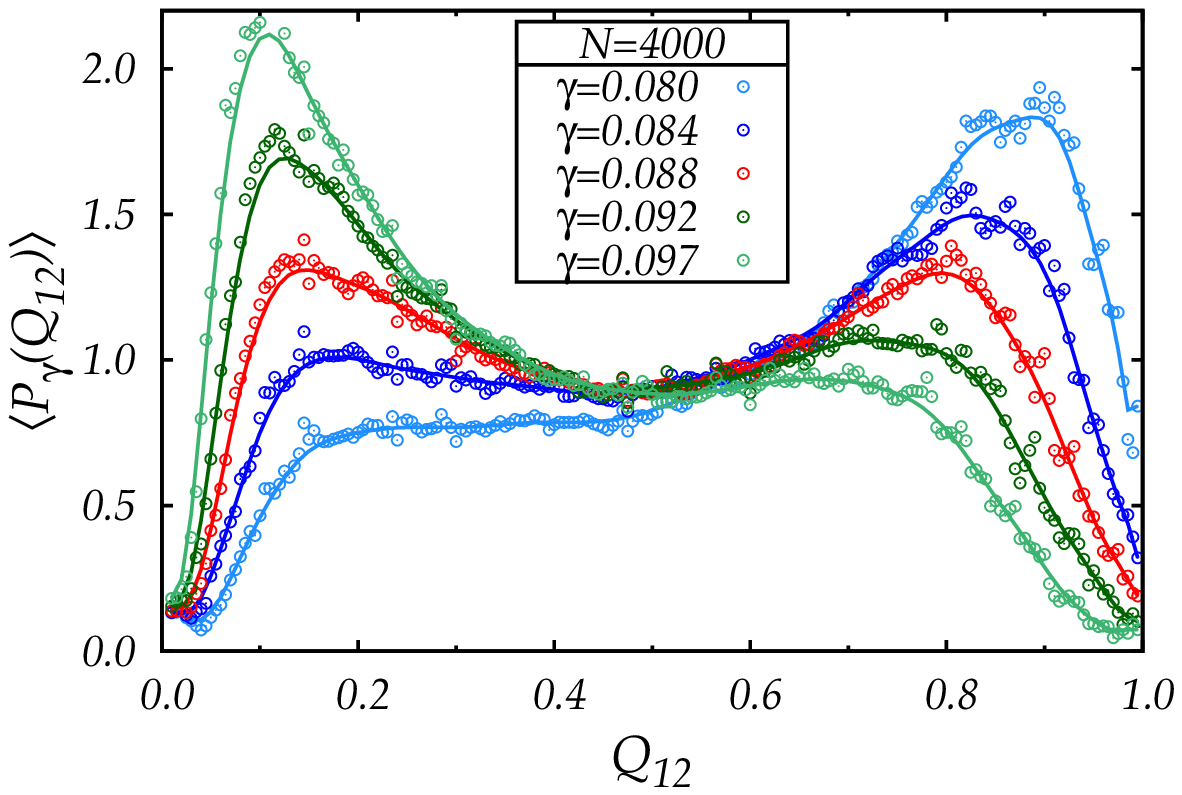}
\caption{Upper panel: the order parameter $\overline{Q_{ab}}$ as a function of the strain $\gamma$,
superimposed on the stress vs. strain curve. Lower panel: the probability distribution function
$P(\overline{Q_{ab}})$ for different values of $\gamma$ in the vicinity of the mechanical yield value $\gamma_{_{Y}}$.}
\label{phasetran}
\end{center}
\end{figure}
In the upper panel of Fig. ~\ref{phasetran} one sees the order parameter $\overline{Q_{ab}}$ as a function
of $\gamma$, superimposed on the stress vs. strain curve of the system under study. The point ``yield" was
obtained with the help of the results shown in the lower panel, in which the probability of observing
$\overline{Q_{ab}}$ is plotted for values of $\gamma$ around the mechanical yield point  $\gamma_{_{Y}}$. The
yield itself is identified when the probability distribution function has two peaks of the same height. The spinodal point is at a slightly higher value of $\gamma$, where the peak occurring around high values
of $\overline{Q_{ab}}$ is about to disappear, with a characteristic spinodal vanishing of the slope. This
is occurring in the present system around $\gamma=0.1$. Of course in the thermodynamic limit the whole
range of $\gamma$ values where the exchange of stability is occurring is becoming very narrow.

It is important to stress again that the ability to observe the divergence of the susceptibility and the
correlation length due to the spinodal phenomenon stems from the fact that we deal with an athermal glassy system
whose typical relaxation times are immense. In a liquid system the fluctuations would have caused the system to make the transition before the spinodal point is reached.

\section*{Conclusions}
In conclusion, we have presented evidence that the yielding transition is a spinodal point with disorder, characterized by a criticality whose features can be picked up by suitable multi-point correlators whose expression can be obtained from replica theory. The treatment presented here pertains to an athermal setting, but an obvious direction for future research will be the application of these ideas to thermal glasses under shear~\cite{16SCH}; in that case, the system will generally be able to escape through thermal activation from the high-$Q_{ab}$ minimum before this has a chance to flatten and the relative susceptibility to diverge. However, since the nucleation time will anyway be fairly long, one should anyway be able to observe \emph{transient} shear-bands/heterogeneities, as long as the temperature is low enough that nucleation does not take place until the system is close to the spinodal, which, interestingly, is precisely the behavior of transient shear bands as reported in~\cite{16SCH}. In this thermal setting, we expect that the study of the ideas presented in this paper will have to proceed much as it does in the case of dynamical heterogeneities around the MCT crossover, entailing for example the definition and study of time-dependent multi-point susceptibilities and correlators.

\acknow{IP acknowledges with thanks receiving the ``Premio Rita Levi-Montalcini" which facilitated the collaboration with GP. We thank George Hentschel and Francesco Zamponi for inspiring discussions. GP acknowldeges funding from the European Research Council (ERC) under the European Union’s Horizon 2020 research and innovation programme (grant agreement No [694925]). IP was supported in part by the Minerva foundation with funding from the Federal German Ministry for Education and Research, and by the Israel Science Foundation (Israel Singapore Program).}

\showacknow % Display the acknowledgments section

%\pnasbreak splits and balances the columns before the references.
% If you see unexpected formatting errors, try commenting out this line
% as it can run into problems with floats and footnotes on the final page.
%\pnasbreak
%\clearpage

% Bibliography
\bibliography{LJ_IT}

\begin{thebibliography}{10}

\bibitem{59LL}
Landau LD, Lifshitz EM (1959) {\em Course of Theoretical Physics Vol 7: Theory
  of Elasticity}.
\newblock (Pergamon Press).

\bibitem{04VBB}
Varnik F, Bocquet L, Barrat JL (2004) A study of the static yield stress in a
  binary lennard-jones glass.
\newblock {\em The Journal of chemical physics} 120(6):2788--2801.

\bibitem{04ML}
Maloney C, Lema\^{i}tre A (2004) Subextensive scaling in the athermal,
  quasistatic limit of amorphous matter in plastic shear flow.
\newblock {\em Phys. Rev. Lett.} 93(1):016001.

\bibitem{05DA}
Demkowicz MJ, Argon AS (2005) Liquidlike atomic environments act as plasticity
  carriers in amorphous silicon.
\newblock {\em Physical Review B} 72(24):245205.

\bibitem{06TLB}
Tanguy A, Leonforte F, Barrat JL (2006) Plastic response of a 2d lennard-jones
  amorphous solid: Detailed analysis of the local rearrangements at very slow
  strain rate.
\newblock {\em The European Physical Journal E} 20(3):355--364.

\bibitem{06LM}
Lema{\^\i}tre A, Maloney C (2006) Sum rules for the quasi-static and
  visco-elastic response of disordered solids at zero temperature.
\newblock {\em Journal of statistical physics} 123(2):415--453.

\bibitem{09LP}
Lerner E, Procaccia I (2009) Locality and nonlocality in elastoplastic
  responses of amorphous solids.
\newblock {\em Phys. Rev. E} 79(6):066109.

\bibitem{11RTV}
Rodney D, Tanguy A, Vandembroucq D (2011) Modeling the mechanics of amorphous
  solids at different length scale and time scale.
\newblock {\em Modelling and Simulation in Materials Science and Engineering}
  19(8):083001.

\bibitem{06SLG}
Subhash G, Liu Q, Gao XL (2006) Quasistatic and high strain rate uniaxial
  compressive response of polymeric structural foams.
\newblock {\em International Journal of Impact Engineering} 32(7):1113--1126.

\bibitem{13KTG}
Kara A, Tasdemirci A, Guden M (2013) Modeling quasi-static and high strain rate
  deformation and failure behavior of a ($\pm$45) symmetric e-glass/polyester
  composite under compressive loading.
\newblock {\em Materials \& Design} 49:566--574.

\bibitem{13NSSMM}
Noradila A, Sajuri Z, Syarif J, Miyashita Y, Mutoh Y (2013) Effect of strain
  rates on tensile and work hardening properties for al-zn magnesium alloys in
  {\em IOP Conference Series: Materials Science and Engineering}.
\newblock (IOP Publishing), Vol.{}~46, p. 012031.

\bibitem{11BB}
Berthier L, Biroli G (2011) Theoretical perspective on the glass transition and
  amorphous materials.
\newblock {\em Rev. Mod. Phys.} 83(2):587--645.

\bibitem{12DHP}
Dasgupta R, Hentschel HGE, Procaccia I (2012) Microscopic mechanism of shear
  bands in amorphous solids.
\newblock {\em Phys. Rev. Lett.} 109(25):255502.

\bibitem{06AG}
Ashby M, Greer A (2006) Metallic glasses as structural materials.
\newblock {\em Scripta Materialia} 54(3):321--326.

\bibitem{13DHP}
Dasgupta R, Hentschel HGE, Procaccia I (2013) Yield strain in shear banding
  amorphous solids.
\newblock {\em Phys. Rev. E} 87(2):022810.

\bibitem{13DGMPS}
Dasgupta R, Gendelman O, Mishra P, Procaccia I, Shor CABZ (2013) Shear
  localization in three-dimensional amorphous solids.
\newblock {\em Phys. Rev. E} 88(3):032401.

\bibitem{16JPRS}
Jaiswal PK, Procaccia I, Rainone C, Singh M (2016) Mechanical yield in
  amorphous solids: A first-order phase transition.
\newblock {\em Phys. Rev. Lett.} 116(8):085501.

\bibitem{16RU}
Rainone C, Urbani P (2016) Following the evolution of glassy states under
  external perturbations: the full replica symmetry breaking solution.
\newblock {\em Journal of Statistical Mechanics: Theory and Experiment}
  2016(5):053302.

\bibitem{16UZ}
{Urbani} P, {Zamponi} F (2016) {Shear yielding and shear jamming of dense hard
  sphere glasses}.
\newblock {\em ArXiv e-prints}.

\bibitem{16NBT}
Nandi SK, Biroli G, Tarjus G (2016) Spinodals with disorder: From avalanches in
  random magnets to glassy dynamics.
\newblock {\em Phys. Rev. Lett.} 116(14):145701.

\bibitem{16SCH}
Shrivastav GP, Chaudhuri P, Horbach J (2016) Heterogeneous dynamics during
  yielding of glasses: Effect of aging.
\newblock {\em Journal of Rheology (1978-present)} 60(5):835--847.

\bibitem{17RL}
Regev I, Lookman T (2017) The irreversibility transition in amorphous solids
  under periodic shear in {\em Avalanches in Functional Materials and
  Geophysics}.
\newblock (Springer), pp. 227--259.

\bibitem{LAS16}
{Leishangthem} P, {Parmar} ADS, {Sastry} S (2016) {The yielding transition in
  amorphous solids under oscillatory shear deformation}.
\newblock {\em ArXiv e-prints}.

\bibitem{02Z-J}
Zinn-Justin J (2002) {\em Quantum field theory and critical phenomena}.
\newblock (Oxford University Press, Oxford).

\bibitem{98DKT}
De~Dominicis C, Kondor I, Temesv{\'a}ri T (1998) Beyond the
  sherrington-kirkpatrick model in {\em Spin glasses and random fields}.
\newblock (World Scientific, Singapore), pp. 119--160.

\bibitem{15RUYZ}
Rainone C, Urbani P, Yoshino H, Zamponi F (2015) Following the evolution of
  hard sphere glasses in infinite dimensions under external perturbations:
  Compression and shear strain.
\newblock {\em Phys. Rev. Lett.} 114(1):015701.

\bibitem{14CKPUZ}
Charbonneau P, Kurchan J, Parisi G, Urbani P, Zamponi F (2014) Fractal free
  energy landscapes in structural glasses.
\newblock {\em Nat. Comm.} 5:3725.

\bibitem{14BB1}
Baity-Jesi M, et~al. (2014) The three-dimensional ising spin glass in an
  external magnetic field: the role of the silent majority.
\newblock {\em Journal of Statistical Mechanics: Theory and Experiment}
  2014(5):P05014.

\bibitem{14BB2}
Baity-Jesi M, et~al. (2014) Dynamical transition in the $d=3$ edwards-anderson
  spin glass in an external magnetic field.
\newblock {\em Phys. Rev. E} 89(3):032140.

\bibitem{16BCJPSZ}
Berthier L, et~al. (2016) Growing timescales and lengthscales characterizing
  vibrations of amorphous solids.
\newblock {\em Proceedings of the National Academy of Sciences}
  113(30):8397--8401.

\bibitem{10KLP}
Karmakar S, Lerner E, Procaccia I (2010) Athermal nonlinear elastic constants
  of amorphous solids.
\newblock {\em Physical Review E} 82(2):026105.

\bibitem{15HJPS}
Gendelman O, Jaiswal PK, Procaccia I, Gupta BS, Zylberg J (2015) Shear
  transformation zones: State determined or protocol dependent?
\newblock {\em EPL (Europhysics Letters)} 109(1):16002.

\end{thebibliography}

\end{document}

\title{Supplementary information for: shear bands as manifestation of a criticality in yielding amorphous solids}

\author{Giorgio Parisi$^1$, Itamar Procaccia$^2$, Corrado Rainone$^2$ and Murari Singh$^2$ }
\affiliation{$^1$Dipartimento di Fisica, Sapienza Universit\'a di Roma, INFN, Sezione di Roma I, IPFC -- CNR, Piazzale Aldo Moro 2, I-00185 Roma, Italy\\$^2$Department of Chemical Physics, the Weizmann Institute of Science, Rehovot 76100, Israel}
\maketitle

\section{The longitudinal correlation function}

Let us start from the expression of the free energy of a glass state, prepared by equilibrating a generic glass former down to a glass transition temperature $T_g$ where it can still be equilibrated, and then quenching it out of equilibrium to a given temperature $T<T_g$. Such a free energy was first defined in~\cite{FP95} in the context of spin-glass physics. Its definition in the case of structural glasses, and its computation in the particular case of hard spheres were first discussed in~\cite{RUYZ15}. The definition, in the case of a generic glass former made of $N$ particles is based on comparing two configurations $X^a$ and $X^b$ of the same glass. Here
\begin{equation}
X^a \equiv \{\B r_i^a\}_{i=1}^N \ , \quad X^b \equiv \{\B r^b_i\}_{i=1}^N \ ,
\end{equation}
where the labeling $\B r_i$ refers to the position of the same particle $i$ in the two different configurations.
For a generic interaction potential $V(X)$ the definition of the free energy is
\begin{equation}
f[T,T_g] \equiv -\frac{1}{\beta N} \int dX_0\ \frac{e^{-\beta_g V(X_0)}}{Z_g}\ \log\left[\int dX_1\ e^{\beta V(X_1)}\delta(q^*_r - Q_{01})\right].
\label{eq:FP}
\end{equation}
where $\beta_g=1/(k_B T_g)$, $\beta=1/(k_B T)$ and $q^*_r$ is the value of $q_r \neq 0$ whereupon the free energy attains a local minimum~\cite{RUYZ15}.
The overlap function $Q_{01}$ for any two configuration, say $a$ and $b$ is~\cite{JPRS16}
\begin{equation}
 Q_{ab} = \frac{1}{N}\sum_{i=1}^N \theta(\ell-|\boldsymbol{r}^a_i - \boldsymbol{r}^b_i|).
\end{equation}
Here $\ell$ is a coarse graining parameter (in~\cite{JPRS16}, $\ell
\simeq 0.3$ in Lennard-Jones units).
The idea is to consider the free energy at temperature $T$ of the glass former, which is \emph{constrained} to stay close to an amorphous configuration $X_0$ which is selected from the equilibrium ensemble, using the canonical distribution when the glass is still at equilibrium at $T_g$.

The properties and computation of the free energy \eqref{eq:FP} are discussed extensively in~\cite{RUYZ15,RU16}, so we refer the interested reader to those works. The explicit analytic computation is accomplished in the
mean field approximation. In our paper we use the results far from the mean field limit, but we ascertain that the relevant correlation functions that are fleshed out in the mean field calculation are the relevant ones also in the general case. Of course, critical exponents can differ. In the sequel we sketch how from this mean-field theory in terms of an overlap order parameter $Q_{ab}$ one can extract the definitions of the correlation functions that are expected to show critical behavior.

The outermost integral in the Eq.~\eqref{eq:FP} can be computed with the replica trick,
\begin{equation}
 f[T,T_g] = \lim_{s\to 0} \partial_s \Phi[T,T_g;s],
\end{equation}
where $\Phi$ is defined as
\begin{equation}
 \Phi[T,T_g;s] =  -\frac{1}{\beta N}\log \int dX_0\ dX_1 \cdots dX_s e^{-\beta_g V(X_0)} e^{- \beta V(X_1)}\delta(q-Q_{01})\cdots e^{- \beta V(X^s)}\delta(q-Q_{0s}),
\end{equation}
so we are considering $s$ replicas of the $X$ configuration. In infinite dimensions for the case of hard spheres it was shown~\cite{KPZ12} that the functional defined above can be written as
\begin{equation}
 \Phi = -\frac{1}{\beta N} \int \mathcal{D} Q_{ab}\ e^{-d S(Q_{ab})} \ .
\end{equation}
Here $\mathcal{D}Q_{ab}$ denotes an integration measure over all the distinct $Q_{ab}$s,
\begin{equation}
 \mathcal{D}Q_{ab} \equiv \prod_{a<b}^{0,s} dQ_{ab},
\end{equation}
and $d$ is the number of spatial dimensions. The functional $S(Q_{ab})$ is referred to as the ``replica action". In the mean-field limit $d\to\infty$, the integral above can be computed via the saddle point method~\cite{BenderAdvancedMethods}, which means that one must consider the optimum points in $Q_{ab}$ of the replica action $S(Q_{ab})$. This means that $S(Q_{ab})$ plays the role of a \emph{Gibbs free energy}, i.e. the free energy for fixed order parameter. An illustrative example is the case of a Curie-Weiss model (mean-field ferromagnet) wherein, for the Helmholtz free energy $F$ in zero magnetic field, one has~\cite{replicanotes}
\begin{equation}
 F(h=0,T) = \min_{m}G(m,T)
\end{equation}
where $G(m,T)$ is indeed the Gibbs free energy for fixed magnetization $m$. The minimization equation for $G$ is then the celebrated equation for the spontaneous magnetization
\begin{equation}
 \frac{\partial G}{\partial m} = 0 \Longrightarrow m = \tanh(\beta m)
\end{equation}
and the ferromagnetic phase transition takes place when the paramagnetic, $m=0$ minimum of $G$ flattens and splits in two degenerate minima with $m\neq 0$, which implies that at the critical temperature $\frac{\partial^2 G}{\partial m^2} = 0$. The derivation of the $S(Q_{ab})$ action in the case of mean-field hard spheres can be found in~\cite{KPZ12}.

In the present case the $f[T,T_g]$ plays the role of the Helmholtz free energy $F$ and the $S(Q_{ab})$ of the Gibbs free energy $G$. With this analogy, one can understand how the critical properties of glass states are related to the matrix of second derivatives of the replica action $S(Q_{ab})$,
\begin{equation}
 M_{ab;cd} \equiv \frac{\partial^2 S}{\partial Q_{a<b}\partial Q_{c<d}},\qquad a,b,c,d \in [1,s]
\end{equation}
in the limit $s \to 0$ (we stress that $X_0$ is not involved in this definition). The inverse $G_{ab;cd}$ of the tensor $M$, defined as
\begin{equation}
\sum_{e\neq f} M_{ab;ef}G_{ef;cd} = \frac{\delta_{ac}\delta_{bd} + \delta_{ad}\delta_{bc}}{2}
\end{equation}
is then the covariance matrix of the mean field theory
\begin{equation}
 G_{ab;cd} = \overline{\left<(Q_{ab}-\left<Q_{ab}\right>)(Q_{cd}-\left<Q_{cd}\right>)\right>},
\end{equation}
wherein the angled brackets denote the thermal average restricted to a single glass sample at temperature $T$ (that is over the canonical distribution of the $X_1$ configuration in the \eqref{eq:FP}), and the overbar denotes the average over all possible glass samples selected at $T_g$ (that is over the canonical distribution of the $X_0$ configuration in the \eqref{eq:FP}). This covariance tensor encodes the critical fluctuations of the system near the critical points whereupon the tensor $M_{ab;cd}$ develops a zero mode.

Let us now assume that the glass state under study is a single minimum of the free-energy landscape of the system wherein all replicas from $1$ to $s$ can move ergodically, this means that the replicas are all equivalent and the matrix $Q_{ab}$ must then be invariant by any replica permutation, an hypothesis referred so as replica-symmetric (RS).\\
In~\cite{RUYZ15} it is discussed how this is not true in all cases, i.e. there exist a regime wherein the glass basin undergoes an ergodicity breaking and fractures into sub-basins. Nevertheless, here we stick to the simple RS ansatz. In this case, since the action $S(Q_{ab})$ must in turn be invariant for any replica permutations, the most general form that the Hessian $M$ can take is
\begin{equation}
 M_{ab;cd} = M_1 \left(\frac{\delta_{ac}\delta_{bd} + \delta_{ad}\delta_{bc}}{2}\right) + M_2 \left(\frac{\delta_{ac} +\delta_{bd} + \delta_{ad} +\delta_{bc}}{4}\right) + M_3,
\end{equation}
and the same goes for the covariance matrix $G_{ab;cd}$. This form is completely general as it only pertains to the RS symmetry; then the only model-dependence is in the parameters $M_1$, $M_2$ and $M_3$, which must be computed case by case and are generally dependent on the external parameters like temperature or magnetic field.
The diagonalization of the tensor $M_{ab;cd}$ is an exercise of standard linear algebra and has been already carried out many times, see for example~\cite{CS92,BM79,DK98} and~\cite{Z10} where it is proposed as an exercise. It is found that the tensor $M$ has only three distinct eigenvalues
\begin{eqnarray}
 \lambda_R &=& M_1\\
 \lambda_L &=& M_1 + (s-1)(M_2+sM_3)\\
 \lambda_A &=& M_1 + \frac{s-2}{2}M_2,
\end{eqnarray}
and the same goes for the tensor $G$. Those three eigenvalues (or modes) are called the \emph{replicon}, \emph{longitudinal}, and \emph{anomalous}, respectively~\cite{Z10}.
We are interested in the longitudinal mode (which in the limit $s\to 0$ is degenerate with the anomalous one), which becomes soft at the yielding transition~\cite{RU16,UZ16}. Let us consider the $G$ tensor. Because of replica symmetry, there are only three distinct correlators that one can define, namely
\begin{eqnarray}
 G_{12;12} &=& \frac{G_1}{2} + \frac{G_2}{2} + G_3\\
 G_{12;13} &=& \frac{G_2}{4} + G_3\\
 G_{12;34} &=& G_3
\end{eqnarray}
and in the limit $s\to0$ we know that
\begin{equation}
 \frac{1}{\lambda_L} = G_1-G_2.
\end{equation}
It is then immediate to check that
\begin{eqnarray}
 G_{12;12} -2G_{12;13} + G_{12;34} &=& \frac{G_1}{2} \propto \frac{1}{\lambda_R} \equiv G_R\\
 G_{12;12} -4G_{12;13} + 3G_{12;34} &=& \frac{G_1-G_2}{2} \propto \frac{1}{\lambda_L} \equiv G_L
\end{eqnarray}
which then implies
\begin{equation}
 G_L(\boldsymbol{r}) = 2G_R(\boldsymbol{r}) -\Gamma_2(\boldsymbol{r}),
\end{equation}
with the definitions
\begin{eqnarray}
 G_R(\boldsymbol{r}) &\equiv& \overline{{\left<Q_{ab}(r)Q_{ab}(0)\right>}} - 2\overline{{\left<Q_{ab}(r)Q_{ac}(0)\right>}} + \overline{{\left<Q_{ab}(r)\right>\left<Q_{cd}(0)\right>}} \label{eq:defGR}\\
 \Gamma_2(\boldsymbol{r}) &\equiv& \overline{{\left<Q_{ab}(\boldsymbol{r})Q_{ab}(0)\right>}} -\overline{{\left<Q_{ab}(\boldsymbol{r})\right>\left<Q_{ab}(0)\right>}}, \label{eq:defG2}
\end{eqnarray}
as in the main text. We have used $\Gamma_2 = G_{12;12} - G_{12;34}$ which derives from replica symmetry, as $\left<Q_{12}\right>\left<Q_{12}\right> = \left<Q_{12}Q_{34}\right>$ in the replica-symmetric phase.

Let us now detail how to transform these definitions into quantities that can be measured in simulation. We start by "localizing" the definition of the $Q_{ab}$ overlap in the following way
 \begin{equation}
  Q_{ab}(\B r) \equiv \sum_{i=1}^N\theta(\ell-|\B r_i^a-\B r_i^b|)\delta (\B r -\B r_i^a).
  \label{eq:defQr}
 \end{equation}
 In a thermal simulation the $a$ and $b$ configurations would depend on the time $t$, and so would the $Q_{ab}(r)$, so one would need to perform the in-state thermal average $\left<\bullet\right>$ by considering the equilibrium value of these quantities. In the present paper we focus un athermal solids under quasi-static shear, so we do not have dynamics and the $a$ and $b$ configurations will simply be two distinct minima of the inter-particle potential obtained through the protocol described in the main text, and the thermal average will be the average over this ensemble of configurations which make up a glassy patch.\\
We now apply the definition \eqref{eq:defQr} in the \eqref{eq:defGR}, \eqref{eq:defG2} to construct the correlators. For illustrative purposes we use the $\Gamma_2(\boldsymbol{r})$. We get, omitting the overline to lighten the notation,
\begin{equation}
 \left<(Q_{ab}(\boldsymbol{x})-\left<Q_{ab}(\boldsymbol{x})\right>)(Q_{ab}(\boldsymbol{x}+\boldsymbol{r})-\left<Q_{ab}(\boldsymbol{x}+\boldsymbol{r})\right>)\right> = \sum_{ij}[(u^{ab}_i  -Q_{ab}) (u^{ab}_j - Q_{ab})]\delta (\B r + \boldsymbol{x} -\B r_i^a)\delta (\B x -\B r_j^a),
\end{equation}
with
 \begin{equation}
 u^{ab}_i \equiv \theta(\ell-|\boldsymbol{r}_i^a-\boldsymbol{r}_i^b|),
 \end{equation}
 as in the main text, and we used that $\left<Q_{ab}(x)\right> = Q_{ab}$. Because of translational invariance, the correlator is actually independent of $\boldsymbol{x}$. We can get rid of $\boldsymbol{x}$ by performing an integration over this variable, which, using the $\delta$-functions, gives as a result
\begin{equation}
\sum_{ij}[(u^{ab}_i  -Q_{ab}) (u^{ab}_j - Q_{ab})]\delta (\B r - (\B r_i^a- \boldsymbol{r}_j^a))
\end{equation}
then, following~\cite{BCJPSZ16}, we omit the terms with $i=j$ (which are anyway relevant only for $\boldsymbol{r} = 0$) and we normalize the correlator with the pair distribution function of the glass; we finally obtain
\begin{equation}
\frac{ \sum_{i\neq j}(u^{ab}_i-Q_{ab}) (u^{ab}_j-Q_{ab})\delta(\boldsymbol{r}-(\boldsymbol{r}_{i}^a-\boldsymbol{r}_{j}^a)) }{ \sum_{i\neq j}\delta(\boldsymbol{r}-(\boldsymbol{r}_{i}^a-\boldsymbol{r}_{j}^a)) } \equiv \tilde \Gamma_2(\B r),
\end{equation}
as in the main text. The derivation for the $\tilde G_R(\boldsymbol{x})$ is then an obvious generalization.

\section{Additional numerical results}

\begin{figure}[htpb]
 \includegraphics[width=0.5\textwidth]{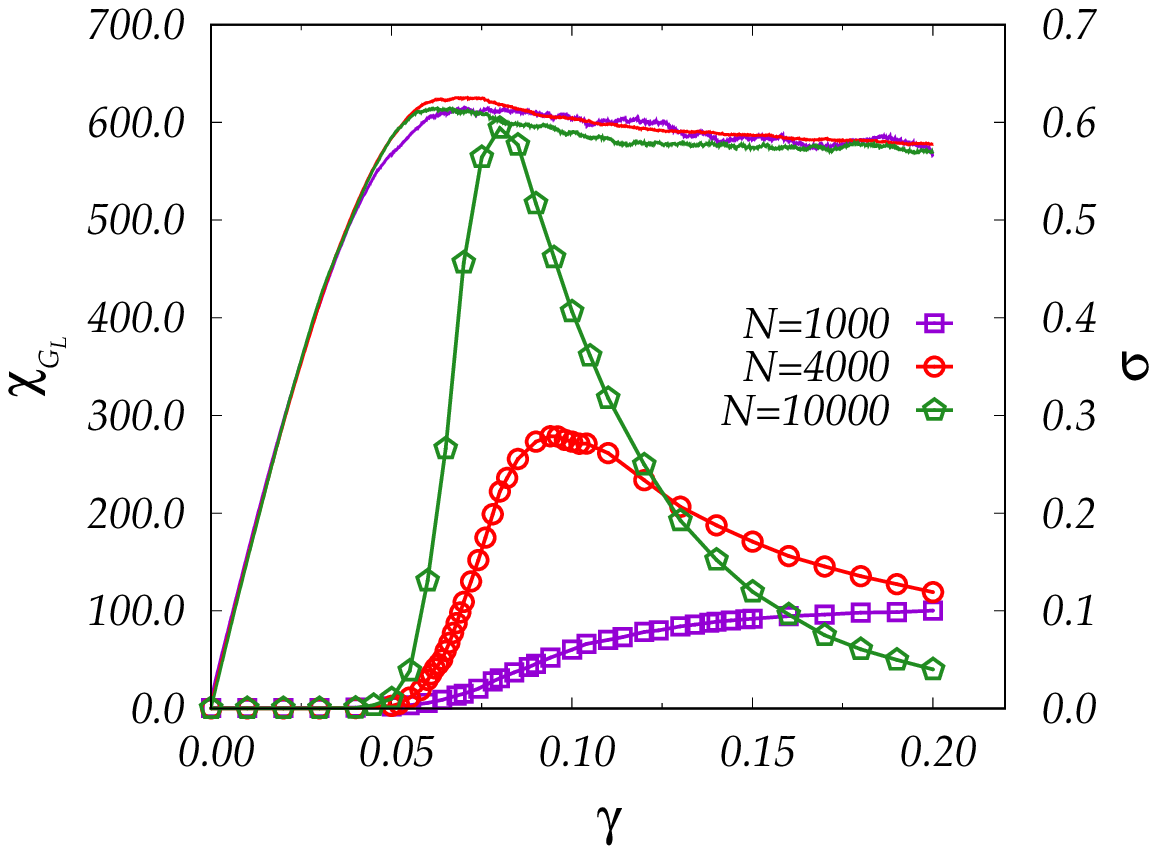}
 \caption{The susceptibility $\chi_{_{G_L}}$  as a function of $\gamma$ for the three systems sizes available. Superimposed are the stress vs. strain curves for comparison.
 The color code is violet for $N=1000$, red for 4000 and green for 10000.}  \label{GL}
\end{figure}

\begin{figure}[htpb]
 \includegraphics[width=0.3\textwidth]{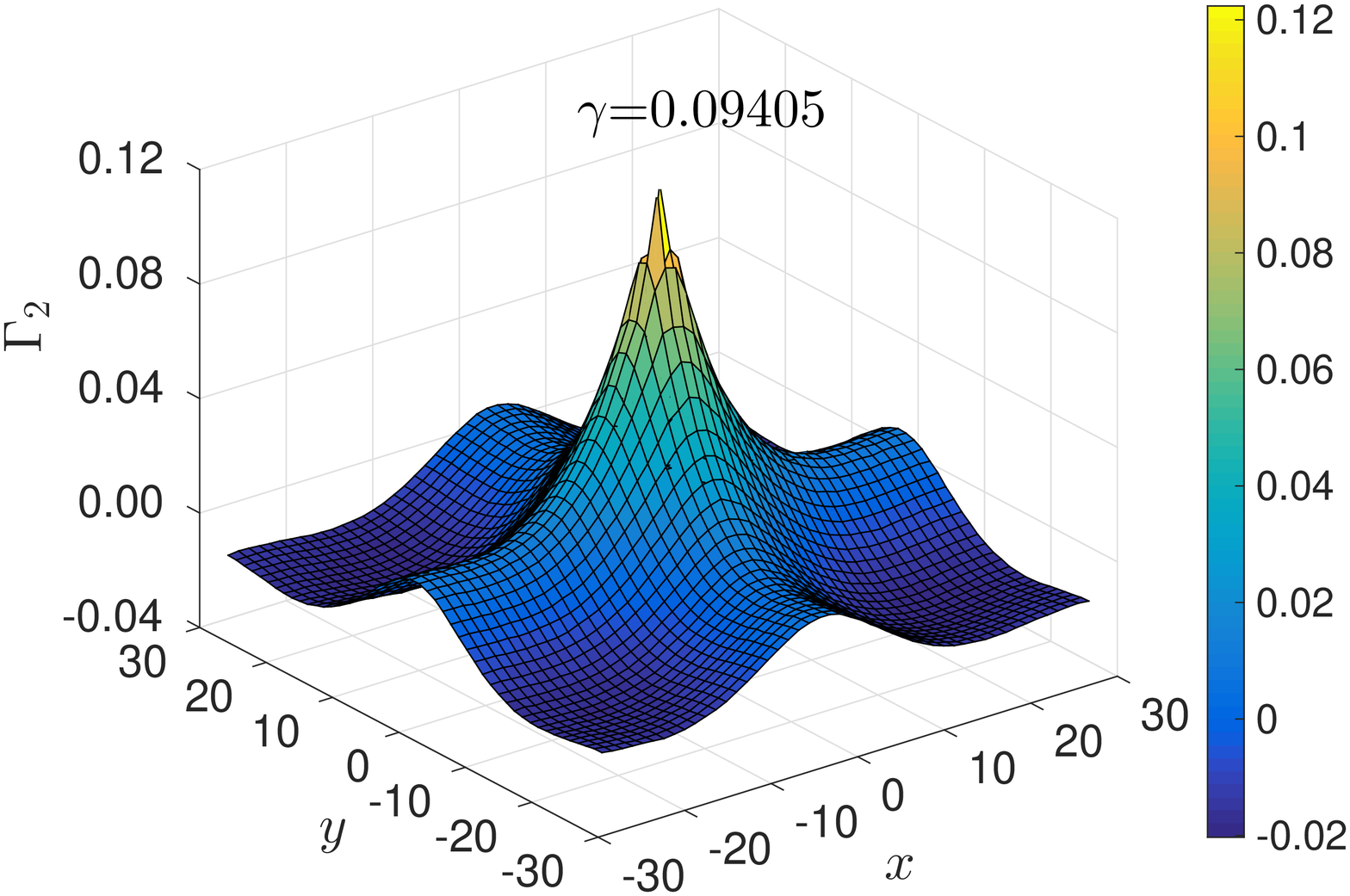}
 \includegraphics[width=0.3\textwidth]{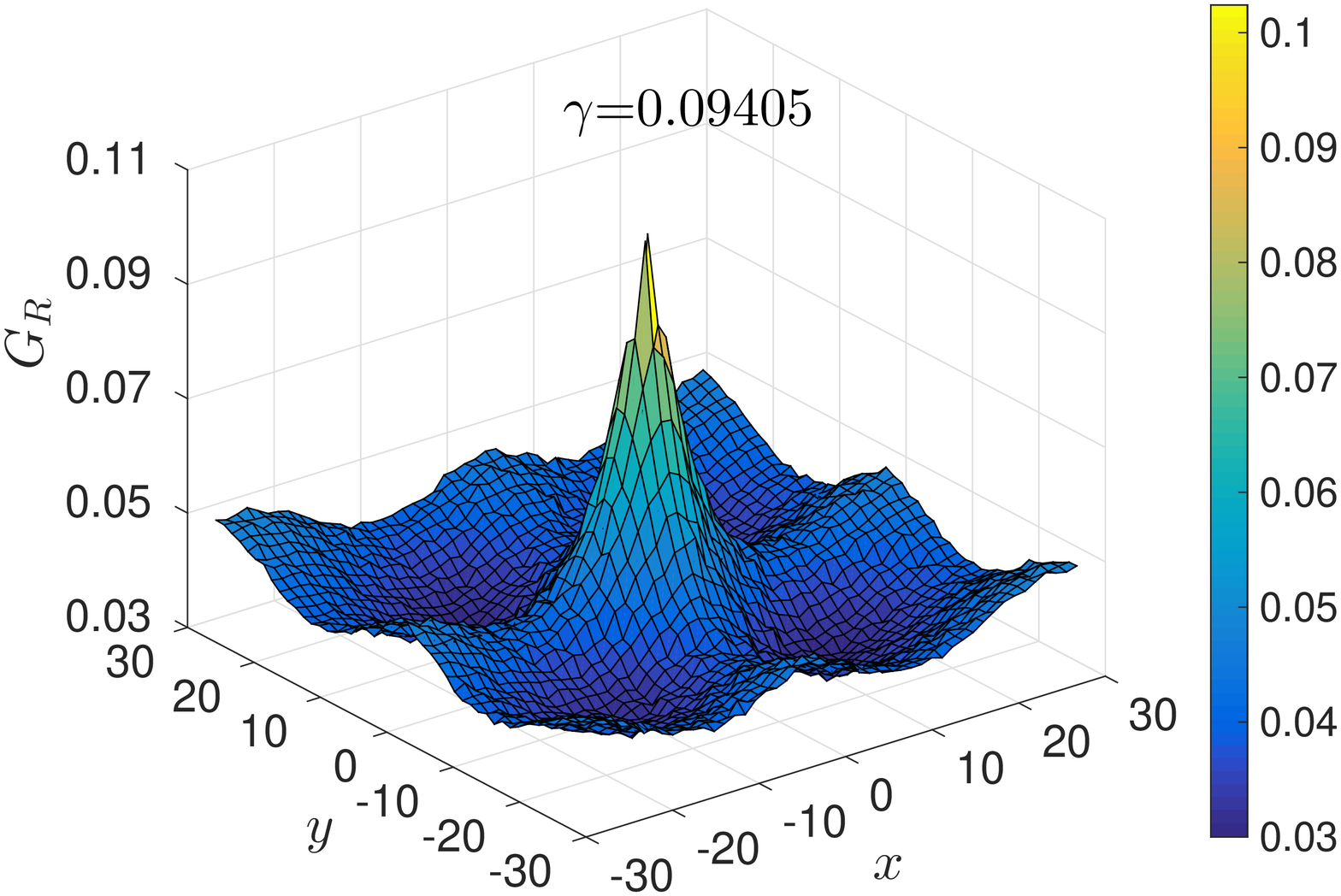}
 \includegraphics[width=0.3\textwidth]{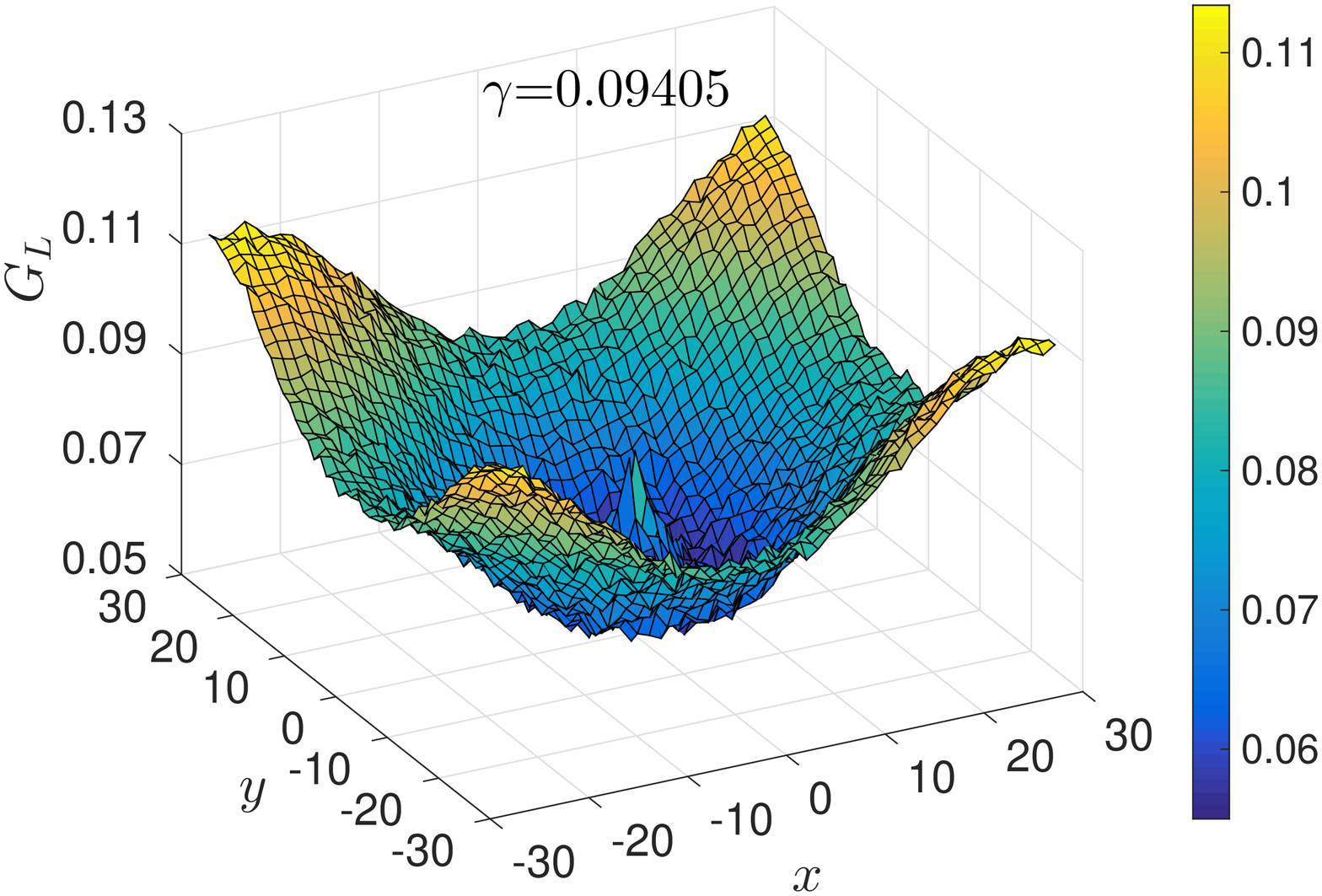}\\
 \caption{A 3-dimensional projection of the three correlation function as a function of $x,y$.}
 \label{correlations}
 \end{figure}

In the paper we have exhibited in Fig. 1 the susceptibilities $\chi_{_{G_R}}$ and $\chi_{_{\Gamma_2}}$. For
completeness we show here also the susceptibility $\chi_{_{G_L}}$, see Fig.~\ref{GL}.
As discussed in the paper, this susceptibility is very close to 2$\chi_{_{G_R}}$. On the other hand
the correlation functions themselves differ $G_R(x,y)$ and $G_L(x,y)$ differ significantly since
the function $\Gamma_2(x,y)$ is not positive definite, becoming negative towards the edges of the
available fields. This is seen clearly in Fig.~\ref{correlations} where we show the three
correlation function at the value of $\gamma=0.09405$.
%%%%%%%%%%%%%%%%%%%%%%%%%%%%%%%%%%%%%%%%%%%%%%%%%%%%%%%%%%%%%%%%%%%%%%
%%%%%%%%%%%%%%%%%%%%%%%%%%%%%%%%%%%%%%%%%%%%%%%%%%%%%%%%%%%%
%%%%%%%%%%%%%%%%%%%%%%%%%%
%%%%%%%%%%%%%%%%%%%%%%%%%%%%%%%%%%%%%%%%%%

%%%%%%%%%%%%%%%%%%%%%%%%%%%%%%%%%%%%%%%%%%%%%%%%%%%%%%%%%%%%%%%%%%%%%%
\begin{figure}[htpb]
 \includegraphics[width=0.45\textwidth]{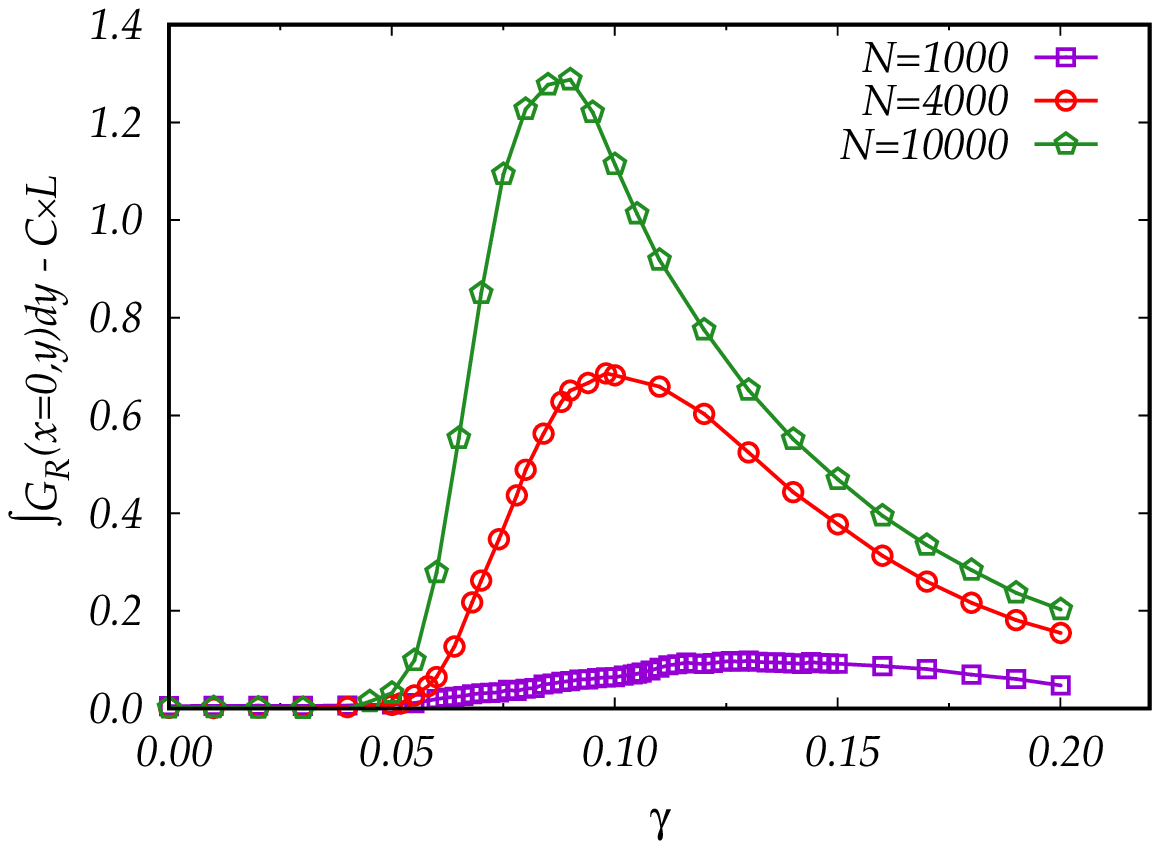}
 \caption{The difference between $\int dy\ G_R(x=0,y)$ and $C\times L$}
\label{Ceffect}
\end{figure}
%%%%%%%%%%%%%%%%%%%%%%%%%%%%%%%%%%%%%%%%%%%%%%%%%%%%%%%%%%%%%%

An interesting observation concerns the constant $C$ used in the fit Eq. 10 in the paper. This constant
is also sensitive to the approach of the criticality, cf. the lower panel in Fig.3 in the paper. One could worry
that integrating this constant over $y$ could contribute to the divergence of the susceptibilities. In fact
the rise in $C$ near the spinodal point goes down with the system size and its contribution to the integral
is reduced as well, as can be seen in Fig.~\ref{Ceffect} which presents the integral $\int dy G_R(x=0,y)$ from
which $C\times L$ is subtracted.
The conclusion is that indeed the contribution of $C$ goes down also when integrated over the system
size, showing that the main contribution to the divergence of the susceptibility is from the divergence
of the correlation length.

\begin{figure}[htb]
 \includegraphics[width=0.45\textwidth]{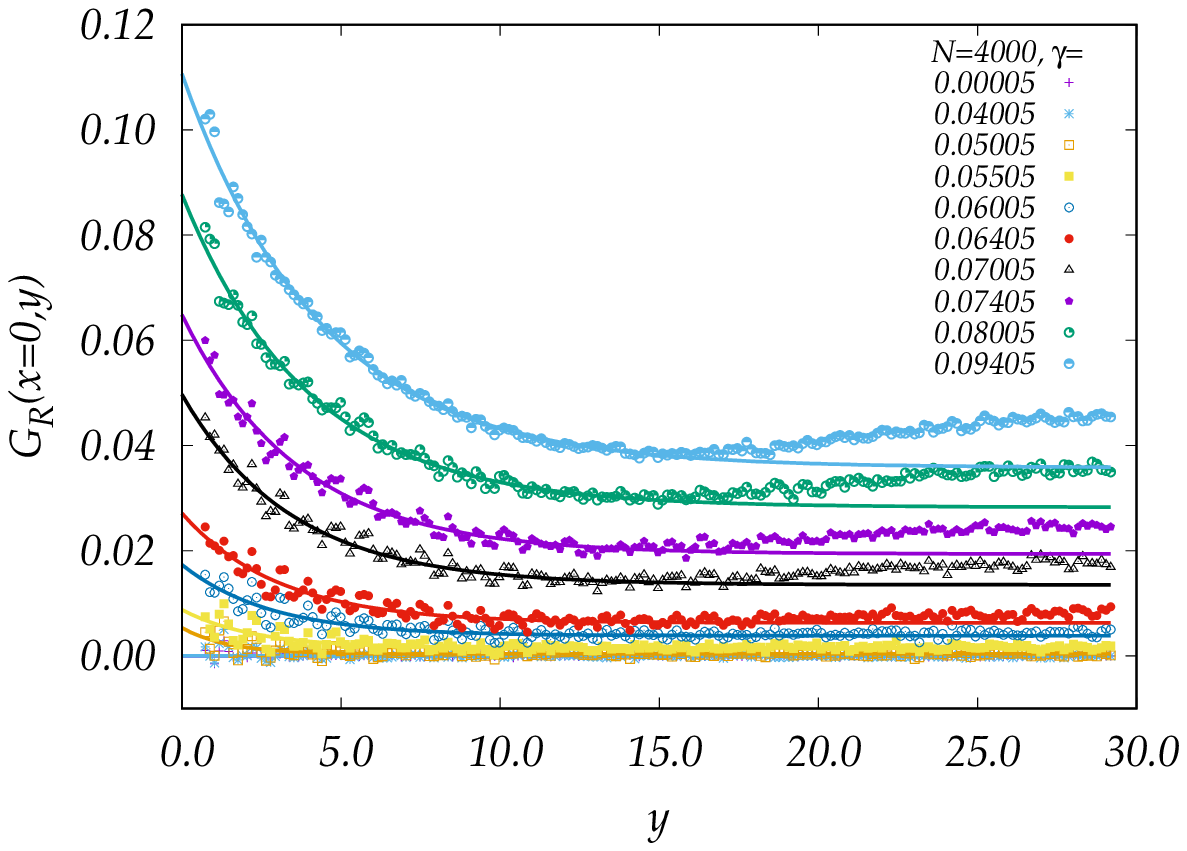}
 \includegraphics[width=0.45\textwidth]{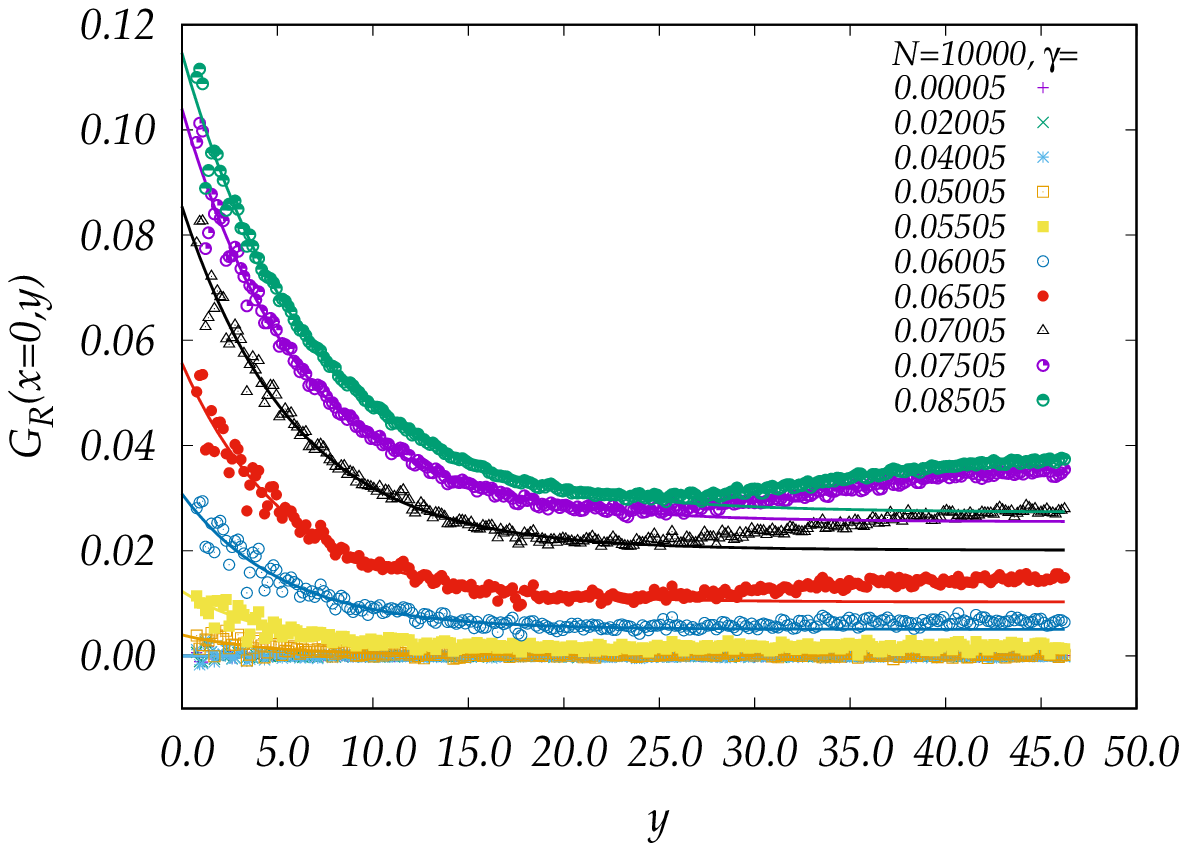}
\caption{The full $y$ dependence of $G_R(x=0,y)$. The region fitted by Eq.~(10) in the paper
is shown. Note that the minimum in the function resides in higher values of $y$ for larger
system sizes.}
\label{full}
\end{figure}

Finally we need to discuss the fitting procedure for the correlation function $G_R(x=0,y)$. In Fig.~\ref{full}
we show the full results for this correlation function for all the available values of $\gamma$ and for two larger systems sizes at our disposal. One sees that the exponents decay that is used for the fit is only
reliable up to the minima of the functions. The reason for the upward trend is the periodic boundary condition that reflects the correlations. To eliminate this spurious effect we presented in the paper the fit
up to the minimum in the function. One should note however that the distance to the minimum increases
with the system size, presumably diverging in the thermodynamic limit. Thus the fit up to the minimum
allows a faithful estimate of the correlation length $\xi$.

%\bibliography{LJ}